\documentclass[5p,twocolumn]{elsarticle}
\usepackage{graphicx,latexsym}
\usepackage{dcolumn}
\usepackage{amssymb,amsmath,bm}
\usepackage{subfigure}
\usepackage{braket}
\usepackage{siunitx}
\usepackage{array}
\usepackage{float}
\usepackage[nopar]{lipsum}
\usepackage{caption}
\usepackage{multirow}
\usepackage{bigstrut}
\usepackage[super]{nth}

\newcommand{\angstrom}{\text{\normalfont\AA}}
\usepackage{hyperref}
\hypersetup{
    pdfnewwindow=true,       
    colorlinks=false,         
    linkcolor=black,          
    citecolor=black,          
    filecolor=magenta,       
    urlcolor=black           
}

\usepackage[normalem]{ulem}

\def\sec#1{Sec.\ \ref{#1}}
\def\eq#1{Eq.\ (\ref{#1})}
\def\fig#1{Fig.\ \ref{#1}}

\journal{}

\begin{document}

\begin{frontmatter}


\title{Properties of BC$_6$N monolayer derived by first-principle computation: 
	   Influences of interactions between dopant atoms}

\author[a1,a2]{Nzar Rauf Abdullah}
\ead{nzar.r.abdullah@gmail.com}
\address[a1]{Division of Computational Nanoscience, Physics Department, College of Science, 
             University of Sulaimani, Sulaimani 46001, Iraq}
\address[a2]{Computer Engineering Department, College of Engineering, Komar University of Science and Technology, Sulaimani 46001, Iraq}

\author[a3]{Botan Jawdat Abdullah}
\address[a3]{Department of Physics, College of Science, Salahaddin University-Erbil, Erbil, Kurdistan Region, Iraq}

\author[a4]{Chi-Shung Tang}
\address[a4]{Department of Mechanical Engineering,
	National United University, 1, Lienda, Miaoli 36003, Taiwan}

\author[a5]{Vidar Gudmundsson}
\address[a5]{Science Institute, University of Iceland, Dunhaga 3, IS-107 Reykjavik, Iceland}


\begin{abstract}

The properties of graphene-like BC$_6$N semiconductor are studied using density functional theory taking into account the attractive interaction between B and N atoms. 	
In the presence of a strong attractive interaction between B and N dopant atoms, the electron charge distribution is highly localized along the B-N bonds, while for a weaker attractive interaction the electrons are delocalized along the entire hexagonal ring of BC$_6$N. 
Furthermore, when both B and N atoms are doped at the same site of the hexagon, the breaking of the sub-lattice symmetry is low producing a small bandgap. In contrast, if the dopant atoms are at different sites, a high sub-lattice symmetry breaking is found leading to a large bandgap.  
The influences of electron localization/delocalization and the tunable bandgap on thermal behaviors such as the electronic thermal conductivity, the Seebeck coefficient, and the figure of merit, and optical properties such as the dielectric function, the excitation spectra, the refractive index, the electron energy loss spectra, the reflectivity, and the optical conductivity are presented.
An enhancement with a red shift of the optical conductivity at low energy range is seen while a reduction at the high energy range is found indicating that the BC$_6$N structure may be useful for optoelectronic devices in the low energy, visible range.

\end{abstract}


\end{frontmatter}

\section{Introduction} 

The optical behavior of 2D materials such as graphene and silicene have unique tunability through controlling the geometrical configuration, the local excitonic effects, the doping density, and the defects levels \cite{doi:10.1063/1.4953172,  C6CP07455C}.
Graphene possesses unique optical transitions and can absorb light over a wide range of frequencies due to the presence of the linear dispersion at the Dirac point.
It leads to an exhibition of extraordinary properties such as a constant optical conductivity 
in the infrared regime and the gate-dependent optical 
absorbance \cite{Wang206, Nair1308, Fan2018}.
The strong optical transitions of a monolayer graphene and the tunability by simple electrical gating hold a promise for new applications in for infrared optics, optoelectronics,
and the semiconductor and optical industry \cite{Romagnoli2018, Jia2012}. 

Substitutional doping has been considered as an effective method for
tailoring the electronic and optical properties of monolayer graphene nanostructures. This is because the doping helps to tune the work function and the carrier concentration of the nanostructure broadening the range of possible electronic and optical applications \cite{Ba2017, doi:10.1021/nl304351z}. Boron (B) atoms supply an electron to the graphene in the case of a B-doped graphene increasing the Fermi energy due to the low electronegativity of the B atom leading to a p-type semiconductor \cite{C5TA10599D}. In contrast, the N atoms receive electrons from the graphene leading to an n-type semiconductor because the electronegativity of an N atom is higher than that of the carbon, C, atoms \cite{Zhang2016}. The optical response of both doped systems is sensitive to their electronic structure around the Fermi energy, and the absorption spectrum undergoes changes in low energy range \cite{GOUDARZI2019130, abdullah2021conversion}.
In the case of BN-codoped monolayer graphene, the optical response evolves from the infrared
to the ultraviolet region and shows significant anisotropy for different polarizations \cite{Bu2020, ABDULLAH2021110095}. Extra peaks in absorption spectra in addition to the $\pi\text{-}\pi^*$ transition peak have been found for BN-codoped graphene in the visible range of the electromagnetic spectrum in the case of parallel polarization \cite{MORTAZAVI2019733}.
Furthermore, the optical response such as the dielectric function and the reflectivity have been investigated by varying the concentration ratio of the BN-dopant atoms in monolayer graphene, and significant red shift in absorption towards the visible range of the radiation at high doping was found \cite{RANI201428}. 

We believe that information on the influences of the interaction between the B and the N atoms in a graphene structure on the optical response is still lacking. We therefore consider the role of the interaction between the B and N dopant atoms in different configuration in a BN-codoped monolayer graphene. 
We find that the imaginary and real part of the dielectric function, the excitation spectra, the refractive index, the reflectivity, and the optical conductivity are prominently modified due to the interaction between the B and N atoms in the system.

In \sec{Sec:Model} the BN-codoped monolayer graphene structure is briefly over-viewed. In \sec{Sec:Results} the main achieved results are analyzed. In \sec{Sec:Conclusion} the conclusion of the results is presented.

\section{Computational details}\label{Sec:Model}

In the present work, the Quantum espresso (QE) package code based on density functional theory (DFT) has been used \cite{Giannozzi_2009, giannozzi2017advanced}. The approach is based on an iterative solution of the Kohn-Sham theory of DFT in a plane-wave set with Norm Conserving (NC) pseudopotential \cite{PhysRev.140.A1133, Petersen2000}. In the calculations, we consider the Perdew-Burke-Ernzerhof (PBE) functional of the GGA approach \cite{PhysRevLett.77.3865, PhysRevB.33.8800, PhysRevB.23.5048}.

Our model is a $2\times 2$ supercell of monolayer sheets, where the sheets are separated by 
20 $\angstrom$ along the perpendicular direction to avoid
an interaction between the layers. The plane wave cutoff energy is assumed to be $680$~eV, and the  Monkhorst-Pack scheme is used to form the Brillouin zone sampling. Our pure and BN-codoped monolayer graphene are fully relaxed with a $8\times8\times1$ $k$-mesh, and the Hellmann-Feynman forces are less than $10^{-4}$ eV/$\angstrom$ per atoms.
The self-consistent field (SCF) calculations have been done with an $8\times8\times1$ $k$-mesh, and a very dense mesh of $k$-points, $100\times100\times1$, is used to calculate the electron density of states (DOS) of the system.

The optical properties of the systems are evaluated using the QE package, and a large number of empty bands is taken into account to evaluate the dielectric properties of the systems, $\varepsilon(\omega) = \varepsilon_1(\omega) + i\varepsilon_2(\omega)$, where $\varepsilon_1$ and $\varepsilon_2$ are the real and imaginary parts of dielectric function. The QE code gives us $\varepsilon_1(\omega)$ and $\varepsilon_2(\omega)$. One can then calculate the complex refractive index, $N(\omega) = n(\omega) + i\, \kappa (\omega)$, where $n(\omega)$ is the real part of refractive index and, $\kappa (\omega)$ the excitation coefficient, the imaginary part of $N$ via \cite{Mistrik2017}

\begin{equation}
	n(\omega) = \frac{1}{\sqrt{2}} \Bigg(  \Big[ \varepsilon_1^2(\omega) + \varepsilon_2^2(\omega)    \Big]^{\frac{1}{2}} + \varepsilon_1(\omega) \Bigg)^{\frac{1}{2}},
	\label{eq_n}
\end{equation}

and 

\begin{equation}
	\kappa (\omega) = \frac{1}{\sqrt{2}}\Bigg( \Big[\varepsilon_1^2(\omega) + \varepsilon_2^2(\omega)    \Big]^{\frac{1}{2}} - \varepsilon_1(\omega) \Bigg)^{\frac{1}{2}}.
	\label{eq_k}
\end{equation}

The reflectivity at normal incidence of EM wave on the materials can be obtained from 
$n$ and $\kappa$ by 

\begin{equation}
	R(\omega) = \frac{\big[n(\omega)-1 \big]^2 + \kappa^2(\omega)}{\big[n(\omega)+2\big]^2 + \kappa^2(\omega)}.
	\label{eq_R}
\end{equation} 

In addition, the optical conductivity can be computed from 

\begin{equation}
	\sigma_{\rm optical} = \frac{-i \, \omega}{4 \pi} \Big[  \varepsilon(\omega) - 1  \Big].
    \label{eq_sigma}
\end{equation}

\section{Results}\label{Sec:Results}

In this section, we present the model, dispersion energy, partial density of states, optical response, and thermal properties of the systems. 

\subsection{Model, Charge distribution, and Interaction energy} 
 
\begin{figure}[htb]
	\centering
	\includegraphics[width=0.45\textwidth]{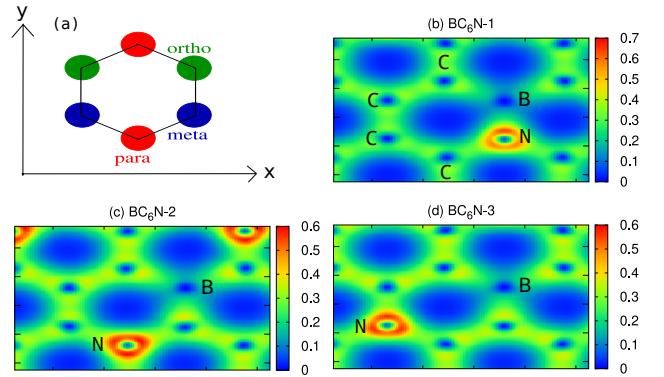}
	\caption{Schematic of a hexagonal structure (a) with the ortho-, the meta-, and the para-positions, and the electron charge density for BC$_6$N-1 (b), BC$_6$N-2 (c), BC$_6$N-3 (d).}
	\label{fig01}
\end{figure}

We consider three different configurations of BN-codoped graphene based on the distance between the B and N atoms. The electron charge distribution of all the three different BN-codped graphenes are shown in \fig{fig01}. 
A schematic diagram of a hexagonal ring is presented in \fig{fig01}(a) indicating the ortho-, the meta-, and the para-position \cite{ABDULLAH2020103282, ABDULLAH2020114556}.
In the first configuration for our systems, the B atom is doped at an ortho-position and the N atom is at a meta-position identified as BC$_6$N-1, and it's electron charge distribution is presented in (b). 
%
The other two configurations are BC$_6$N-2 (c) and BC$_6$N-3 (d) where the B atom is fixed at an ortho-position, but the N atom is moved to a para-, or a meta-position, respectively. 
In this way, we define three BC$_6$N structures with different distances and interaction strength between the B and the N dopant atoms. The distance between the B and the N atom in BC$_6$N-1, BC$_6$N-2, and BC$_6$N-3 is $1.41$, $2.43$, and $2.8\, \angstrom$, respectively. The interaction energy between the B and the N atoms can be deduced from the total energy obtained via a SCF calculation \cite{doi:10.1063/1.4742063, ABDULLAH2020100740}.
The interaction type between the B and the N atoms is attractive as the interaction energy has a negative sign \cite{ABDULLAH2020100740}. Our calculations indicate that the strongest interaction between the B and the N atoms is in BC$_6$N-1, which is expected as the distance between the B and the N atoms is then the shortest for all the three configurations. The attractive interaction becomes weaker increasing distance between the B and the N atoms in both BC$_6$N-2, and BC$_6$N-3.

The C, B and N atoms can be ordered from the higher to lower electronegativity as follows: N $>$ C $>$ B. The B atom gives an electron to the C atoms, and the N atom receives electrons from the C atoms due the difference in their electronegativity \cite{ABDULLAH2020126350}. Consequently, one can see that the electron charge distribution around the N atom is high and around the B atom is low comparing to the C atoms in all three configuration structures shown in \fig{fig01}. 
In BC$_6$N-1, electrons are directly transferred from a B atom to an N atom leading to a higher electron charge distribution in the B-N bonds near to the N atom comparing to the C-B and C-N bonds. This confirms the strong attractive interaction between the B and N atoms.
In both BC$_6$N-2 and BC$_6$N-3, electrons are not directly transferred from a B atom to an N atom, but an N atom receives electrons from the C atoms along all the three C-N bonds in the hexagonal, and a C atoms receives electrons from the B atoms. As a results, the electron delocalization along the C-C, C-B and C-N bonds is indicative for electron transport along the entire ring of the hexagonal structure. This gives rise a higher electron charge distribution between the C-C, C-B, and C-N bonds of the BC$_6$N-2 and BC$_6$N-3 structures comparing to the BC$_6$N-1 structure.  

The stability of the structures can be determined by the bonding between the C, B and N atoms, and the number of C-C, C-B, C-N, and B-N bonds in a structure will affect its stability. The binding energies of the C-C and B-N bonds are higher than those of the C-B and C-N bonds. So, a structure with a higher number of C-C and B-N bonds has a higher binding energy \cite{C2NR11728B, ABDULLAH2021114644}. We can therefore confirm that the BC$_6$N-1 structure is the most energetically stable structure. 

\begin{figure}[htb]
	\centering
	\includegraphics[width=0.4\textwidth]{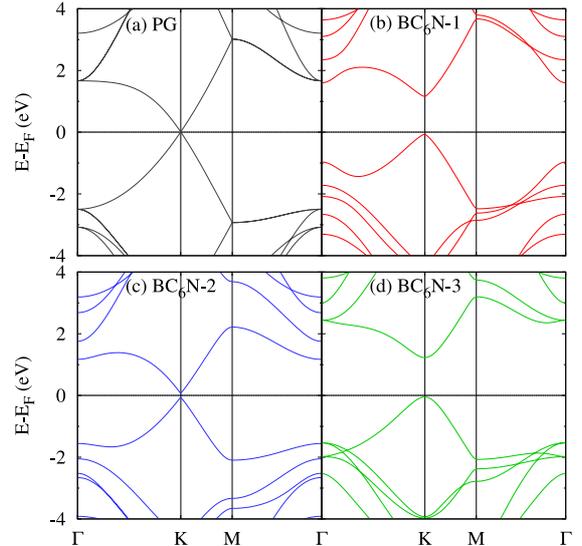}
	\caption{Electronic band structure of PG (a), and BN-codoped graphene with BC$_6$N-1 (b), BC$_6$N-2 (c), and BC$_6$N-3 (d) configurations. The Fermi energy is set to zero \cite{ABDULLAH2020126807}.}
	\label{fig02}
\end{figure}

\subsection{Dispersion energy and PDOS}

The electronic dispersion energy for the BC$_6$N structures (b-d) are displayed in \fig{fig02} together with that of pure graphene (a). The band structure of pure graphene (PG) is re-plotted for the sake of comparing it with the band structure of BC$_6$N. 

A bandgap opens in the presence of the B and N atoms reflecting the sublattice symmetry breaking. 
The bandgap is $0.13$, $1.2$ and $1.26$~eV for BC$_6$N-1, BC$_6$N-2, and BC$_6$N-3, respectively.
In spite of the sublattice symmetry breaking, the Dirac cone of the BC$_6$N structures remains preserved. One notices that the opening of a bandgap is larger for the structures whose B and N atoms are doped at both A- and B-sites of the hexagonal ring such as BC$_6$N-1 and BC$_6$N-3. If the B and N atoms are doped at either a A- or a B-site such as in BC$_6$N-2, the bandgap is small indicating smaller symmetry breaking. 
This has been previously explained and confirmed by Zhu and \emph{et~al.} \cite{doi:10.1021/jp2016616}, and Chang and \emph{et~al.} \cite{doi:10.1021/jp302293p} where they have connected the opening up of a bandgap to the number of $\pi$-bands or $\pi$-electrons, and aromaticity of the structures, respectively. 

\begin{figure}[htb]
	\centering
	\includegraphics[width=0.4\textwidth]{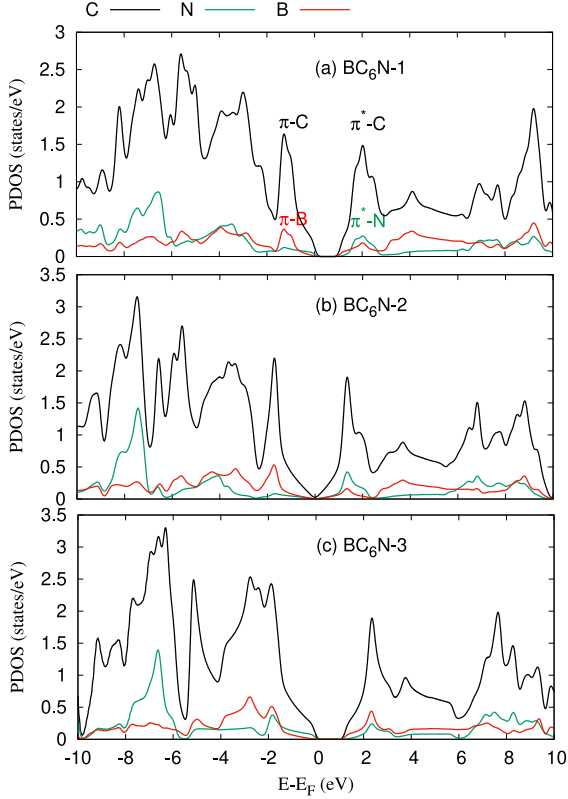}
	\caption{Partial density of state, PDOS, for the BC$_6$N-1 (a), BC$_6$N-2 (b), and BC$_6$N-3 (c) configurations. The $\pi$-C and $\pi^*$-C are the PDOS due to $p_z$-component of the C atoms, and  $\pi$-B and $\pi^*$-N refer to the contribution of $p_z$-components of B and N atoms, respectively. The Fermi energy is set to zero.}
	\label{fig03}
\end{figure}

They show that there is an odd number of $\pi$-electron or $\pi$-bands if the B and N atoms are doped at either a A- or a B-site producing a small bandgap, while an even number of $\pi$-electrons is found for a structure with the B and N atoms doped at a A- and a B-site leading to a larger bandgap.
We draw the attention to that the dispersion energies of these structures are also presented in \cite{ABDULLAH2020126807}, where the mechanical and thermal properties of the systems were investigated.

The partial density of states (PDOS) for BC$_6$N-1 (a), BC$_6$N-2 (b), and BC$_6$N-3 (c)  structures are demonstrated in \fig{fig03}. 
It can be seen from the PDOS that both the B and N atoms contribute to the opening up of a bandgap as the PDOS of the B and N atoms have a high contribution around the Fermi energy, and the maxima and minima of the valence and conduction band edge in the energy range from $-2.5$ to $2.5$~eV.
The PDOS around the Fermi energy for both the B and N atoms are p$_z$-components indicating the $\pi$-states of a B and an N atoms.
Furthermore, the distance between the $\pi$-peak and the $\pi^*$-peak in the PDOS around the Fermi energy for BC$_6$N-1 and BC$_6$N-3 is $3.289$ and $4.17$~eV, respectively. 
These are larger than for BC$_6$N-2, $3.068$~eV, confirming the larger bandgaps of these two structures.

\subsection{Optical properties}

It was seen in the previous section that the B-N dopant atoms can significantly 
tune the dispersion energy and PDOS of BC$_6$N-monolayer structures. 
This modification of the band structure and PDOS alter the optical properties of systems.
In this section, we focus on the main results of the dielectric function, the refractive index, 
the reflectivity, the electron energy loss spectra, EELS, and the optical conductivity of BC$_6$N-monolayer structures emerging due to the B and N dopant atoms. 
In general, the optical response rather depends on the position of the B and N atoms doped at either the same or different sites in the hexagonal structure of graphene.

\begin{figure}[htb]
 	\centering
 	\includegraphics[width=0.4\textwidth]{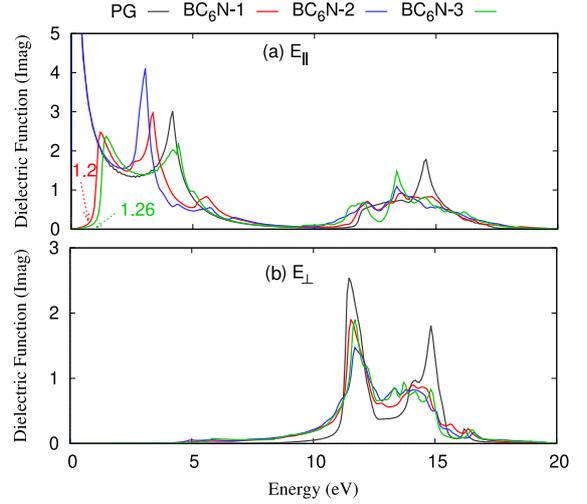}
 	\caption{Imaginary part of dielectric function, $\varepsilon_2$, for PG (black), BC$_6$N-1 (red), BC$_6$N-2 (blue), and BC$_6$N-3 (green) for both parallel, E$_{\parallel}$ (a) and perpendicular, E$_{\perp}$ (b), polarization of the electric field vector of the external EM field.}
 	\label{fig04}
\end{figure}

The imaginary part of dielectric function that expresses the damping of the wave and the dissipation of energy is presented in \fig{fig04}  with both parallel, E$_{\parallel}$ (a) and perpendicular, E$_{\perp}$ (b), polarization of electric field vector of external EM field. 
%
In order to clearly see the modification of the optical response, we re-plot the dielectric function of PG.
In the case of E$_{\parallel}$, two main peaks in imaginary part of dielectric function (black), $\varepsilon_2$, are observed for PG at $\approx 4.16$ and $14.59$~eV corresponding to the $\pi\text{-}\pi^*$, and $\sigma\text{-}\sigma^*$ transitions, respectively. The position of the first peak is close to the experimental value observed at $4.5$~eV \cite{PhysRevB.81.155413}. 
The slight different in the peak position stems from the neglect of the interaction between the monolayer graphene and the substrate in our calculation. 
The position of the second peak agrees very well with an experimental value, $14.6$~eV, obtained for free-standing monolayer graphene \cite{PhysRevB.77.233406}.
In addition to the two peaks, a strong peak at zero energy is seen due to the vanishing 
bandgap of PG.
In the case of E$_{\perp}$ shown in \fig{fig04}(b), two peaks at $11.46$, and $14.83$~eV for PG are seen associated with the $\pi\text{-}\sigma^*$, and $\sigma\text{-}\pi^*$ transitions, respectively.

In the presence of the BN-dopant atoms, the Fermi level shifts and a bandgap is introduced at Fermi level so the energy peak at $4.16$~eV is prominently shifted. In addition to the two main peaks, a pronounced peak starting from $1.2$ (red), or $1.26$~eV (green) is found in the visible region for BC$_6$N-1, and BC$_6$N-3, respectively, in the case of E$_{\parallel}$. These energy values coincide with the bandgap of the two structures, BC$_6$N-1, and BC$_6$N-3. So, the peak in the visible region is caused by the opening up of a bandgap. 
Furthermore, a very strong peak at $0.1$~eV is seen for BC$_6$N-1 indicative of the transition around small bandgap of the structure.
In general, for both E$_{\parallel}$ and E$_{\perp}$, the intensity of both main peaks is decreased for all three BN-codoped graphene structures indicating a weaker damping or dissipation of the wave compared to PG. This reflects the increased localization 
of the electrons occurring in these structures.

\begin{figure}[htb]
	\centering
	\includegraphics[width=0.4\textwidth]{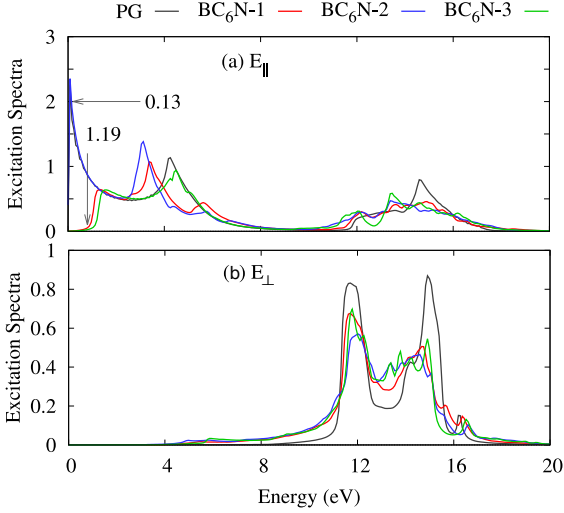}
	\caption{Excitation spectra (imaginary part of refractive index ($\kappa$), for PG (black), BC$_6$N-1 (red), BC$_6$N-2 (blue), and BC$_6$N-3 (green) for both parallel, E$_{\parallel}$ (a) and perpendicular, E$_{\perp}$ (b), polarization of electric field vector of external EM field.}
	\label{fig05}
\end{figure}

The imaginary part of a refractive index (excitation spectra) shown in \eq{eq_k}, $\kappa (\omega)$, is related to the absorption coefficient. The excitation spectra is qualitatively very similar to the imaginary part of the dielectric function for both E$_{\parallel}$ (a) and E$_{\perp}$ (b) displayed in \fig{fig05} for PG and BN-codoped graphene. 
Similar information can be read from $\varepsilon_2$, 
a significant red shift of the two main peaks of the excitation spectra
towards the visible range of the electromagnetic radiation is observed for BN-codoped graphene structures in the case of E$_{\parallel}$, while a reduction in the peaks without any pronounced shift is seen in the case of E$_{\perp}$. A peak caused by the opening up of a bandgap is also seen for E$_{\parallel}$.

\begin{figure}[htb]
	\centering
	\includegraphics[width=0.4\textwidth]{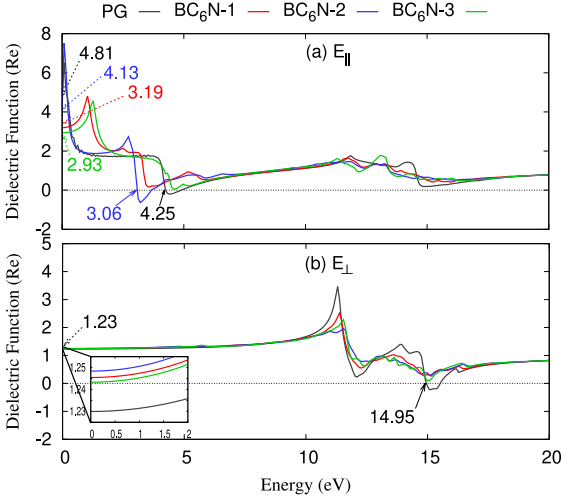}
	\caption{Real part of dielectric function, $\varepsilon_1$, for PG (black), BC$_6$N-1 (red), BC$_6$N-2 (blue), and BC$_6$N-3 (green) for both parallel, E$_{\parallel}$ (a) and perpendicular, E$_{\perp}$ (b), polarization of the electric field vector of an external EM field.}
	\label{fig06}
\end{figure}

\begin{figure}[htb]
	\centering
	\includegraphics[width=0.4\textwidth]{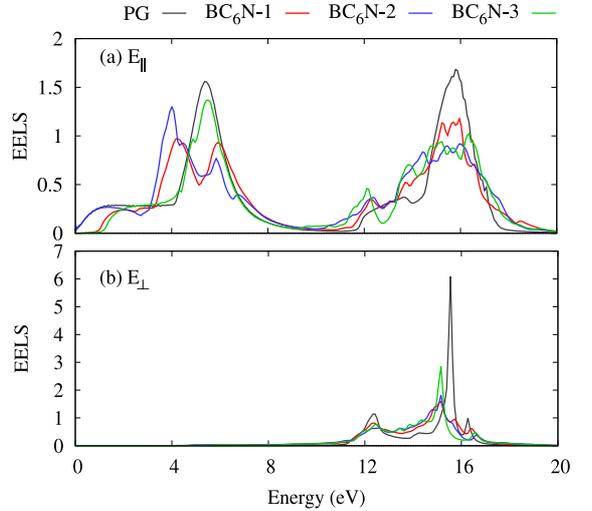}
	\caption{EELS for PG (black), BC$_6$N-1 (red), BC$_6$N-2 (blue), and BC$_6$N-3 (green) for both parallel, E$_{\parallel}$ (a) and perpendicular, E$_{\perp}$ (b), polarization of electric field vector of external EM field.}
	\label{fig07}
\end{figure}

Further interesting information about optical response is associated with the real part of dielectric function, $\varepsilon_1$, which 
is presented in \fig{fig06} for both parallel, E$_{\parallel}$ (a) and perpendicular, E$_{\perp}$ (b), polarization of the external electric field. The $\varepsilon_1$ is indicative of the ability of materials to store electrical energy related to the polarization. 
The value of $\varepsilon_1(0)$ is defined as a static dielectric constant or the value of dielectric function at zero energy.
%
The value of $\varepsilon_1(0)$ of PG is $4.81$, and $1.23$ for E$_{\parallel}$ and E$_{\perp}$, respectively, which is in a good agreement with the recent results of $\varepsilon_1(0)$ for both directions of polarized electric field calculated via DFT \cite{nano8110962}, and in a fair agreement with \cite{RANI201428} for E$_{\parallel}$, where $7.6$ and $1.25$ are obtained. The different values of $\varepsilon_1(0)$ for PG may be influenced by the number of bands taken in the DFT calculations or choice of pseudopotentials. 
When the value of $\varepsilon_1$ is zero or very small, and $\varepsilon_2$ is also small, the electron energy loss spectra (EELS) shown in \fig{fig07} starts to display a peak. 
This frequency is the plasma frequency for collective electron oscillations, $\omega_p$, 
and the energy of plasma frequency is defined as $E = \hbar \omega_p$.  
The energy of the plasmon oscillation for PG is found to be $4.25$ and $14.95$~eV for E$_{\parallel}$ and E$_{\perp}$, respectively, agreeing very well with \cite{NATH2014275}.

In the presence of BN dopant atoms the value of $\varepsilon_1(0)$ does not change appreciably for E$_{\perp}$ polarization. However, for E$_{\parallel}$ polarization, the values of $\varepsilon_1(0)$ and $\varepsilon_1(\omega)$ below $5$~eV change significantly in the presence of B and N atoms doped at the A- and B-sites of the hexagon, BC$_6$N-1, and BC$_6$N-3. This is caused by the stronger symmetry breaking of the sub-lattice symmetry when both the A- and B-sites are doped.
The value of $\varepsilon_1(0)$ is reduced to $3.19$ and $2.39$ for BC$_6$N-1, and BC$_6$N-3, respectively, in the case of E$_{\parallel}$ indicating 
the weaker ability of these structures to store electrical energy at nearly zero-energy of an applied electric field, and it is a slightly increased in the case of E$_{\perp}$ (inset of \fig{fig06}).
On the other hand, for BC$_6$N-2 where both the B and N atoms are doped at the A- or the B-sites resulting in less breaking of the sub-lattice symmetry, the value of $\varepsilon_1(0)$
is found to be $4.13$, which is relatively close to that of PG. We therefore see that the ability of BC$_6$N-2 to store energy is almost similar to PG.
Furthermore, the plasma oscillation is only seen at $3.06$~eV for BC$_6$N-2 in the case of E$_{\parallel}$ while the plasma oscillation vanishes for both BC$_6$N-1, and BC$_6$N-3.

\begin{figure}[htb]
	\centering
	\includegraphics[width=0.4\textwidth]{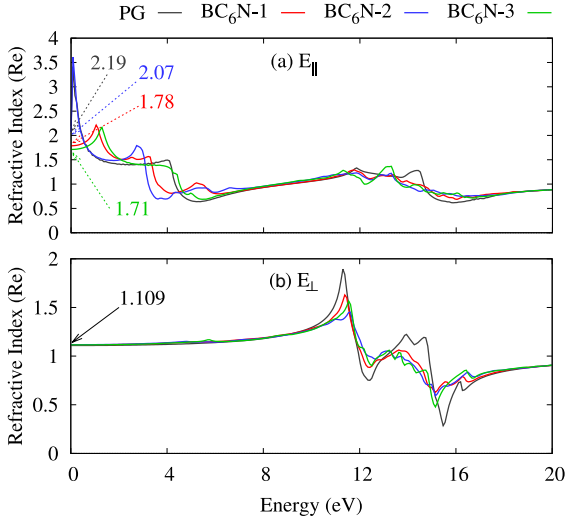}
	\caption{Refractive index, real part of complex refractive index, for PG (black), BC$_6$N-1 (red), BC$_6$N-2 (blue), and BC$_6$N-3 (green) for both parallel, E$_{\parallel}$ (a) and perpendicular, E$_{\perp}$ (b), polarization of electric field vector of external EM field.}
	\label{fig08}
\end{figure}

The real part of the complex refractive index, $n(\omega)$, called the refractive index has been calculated from the real and imaginary parts of dielectric function using \eq{eq_n}. 
The refractive index defines the transmission efficiency of a system.
The refractive index for PG (black), BC$_6$N-1 (red), BC$_6$N-2 (blue), and BC$_6$N-3 (green) for both parallel, E$_{\parallel}$ (a) and perpendicular, E$_{\perp}$ (b), polarization of electric field is presented in \fig{fig08}.
By comparing the $\varepsilon_1$ of \fig{fig06} and the refractive index of \fig{fig08},
we see that change trends of \fig{fig06}(a,b) are similar with those in \fig{fig08}(a,b). This indicates that the influence of $\varepsilon_1$ on $n(\omega)$ plays the leading role.

The real part of refractive index of PG near to zero energy, $n(0)$, is found to be $2.19$ for E$_{\parallel}$ and $1.109$ for E$_{\perp}$. In the presence of the attractive interaction between the B and N atoms, no noticeable change is found
corresponding to the $n(0)$ value the peak positions for the E$_{\perp}$ polarization. 
In the case of E$_{\parallel}$, the value of $n(0)$ is observed to be $1.78$, $2.07$, and $1.71$ indicating that the refractive index is reduced by $18.73\%$, $5.43\%$, and $21.92\%$ for BC$_6$N-1, BC$_6$N-2, and BC$_6$N-3, respectively, compared to PG. The reduction ratio in the refractive index is higher for BC$_6$N-1 and BC$_6$N-3 where both the B and N atoms are doped at different atomic sites indicating less transmittance behaviors near zero-energy.

\begin{figure}[htb]
	\centering
	\includegraphics[width=0.4\textwidth]{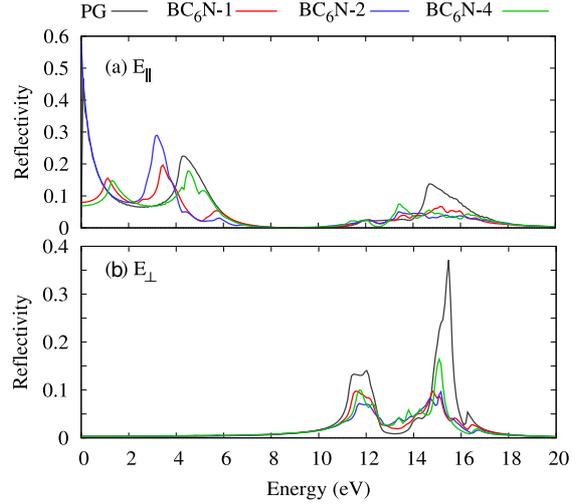}
	\caption{Reflectivity for PG (black), BC$_6$N-1 (red), BC$_6$N-2 (blue), and BC$_6$N-3 (green) for both parallel, E$_{\parallel}$ (a) and perpendicular, E$_{\perp}$ (b), polarization of electric field vector of external EM field.}
	\label{fig09}
\end{figure}

After obtaining the frequency dependent refractive index, we compute
the reflectivity of the PG and BN-codoped graphene for both E$_{\parallel}$ (a) and E$_{\perp}$ (b) polarization shown in \fig{fig09}.    
Two main peaks in the reflectivity of PG are found at 
$4.32$ and $14.76$~eV for E$_{\parallel}$, and two main peaks at $12.02$ and $15.47$~eV for E$_{\perp}$. The peak positions for the reflectivity for PG in both directions of electric field are well agree with \cite{NATH2014275}.
For all three configurations of the BC$_6$N structures, the peak intensity of the reflectivity located between $10$ to $20$~eV for both directions of E-field is reduced irrespective of the strength of the attractive interaction between the B and N atom, and the position of the B-N atoms. The collective oscillations of free electrons and the valence electrons in the selected energy range, $10\text{-}20$~eV, are activated leading to the reduction in reflectivity of the BC$_6$N structures.  
In contrast, an enhancement of the reflectivity is seen for BC$_6$N-2 in the ultra violet (UV) region at a frequency of $<6.0$~eV for E$_{\parallel}$. This is in agreement with our previous explanation of the electron charge density distribution shown in \fig{fig01}, where the electron delocalization in BC$_6$N-2 is high leading to higher reflectivity in the UV region. 
The reflectivity for PG and all the BN-codoped structures is very small or close to zero in the energy range $7\text{-}10$~eV for both directions of E-field while 
the refractive index is close to unity in this energy range.

\begin{figure}[htb]
	\centering
	\includegraphics[width=0.4\textwidth]{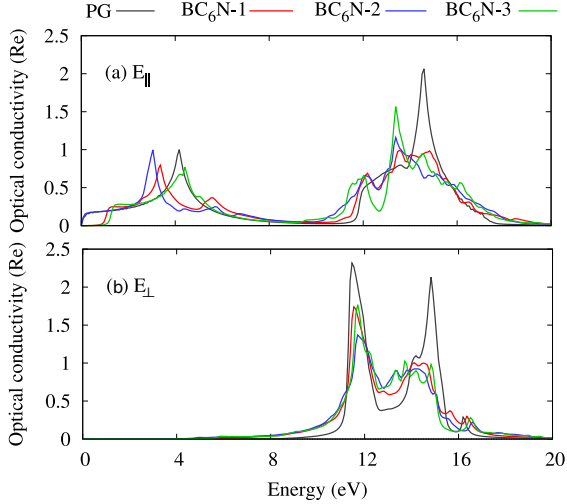}
	\caption{Optical conductivity, $\text{Re}(\sigma_{\rm optical})$ for PG (black), BC$_6$N-1 (red), BC$_6$N-2 (blue), and BC$_6$N-3 (green) for both parallel, E$_{\parallel}$ (a) and perpendicular, E$_{\perp}$ (b), polarization of electric field vector of external EM field.}
	\label{fig10}
\end{figure}

Last, the real part of the optical conductivity is shown in \fig{fig10} for both a parallel, E$_{\parallel}$ (a) and a perpendicular, E$_{\perp}$ (b) electric field. 
The optical conductivities starts with a gap in the case of E$_{\parallel}$, which is due to the semiconducting properties of the BC$_6$N-1 and BC$_6$N-3. A very small zero value of optical conductivity of BC$_6$N-2 is seen in the low energy range $<0.15$~eV indicative of the small bandgap of the structure. 
For PG, two main peaks in the optical conductivity are seen at $4.33$ and $14.75$~eV for E$_{\parallel}$ indicating the $\pi\text{-}\pi^*$ and $\sigma\text{-}\sigma^*$ transitions, respectively, and $12.02$ and $15.47$~eV for E$_{\perp}$ showing $\pi\text{-}\sigma^*$ and $\sigma\text{-}\pi^*$, respectively.
%
As we have mentioned, the peak intensity of the optical conductivity located between $10$ to $20$~eV for both directions of an E-field is reduced irrespective of the strength of attractive interaction between the B and N atom, and the position of B-N atoms. 
In contrast, the peak position 
in the energy range of $3\text{-}6$~eV is shifted towards a lower energy for BC$_6$N-1 and BC$_6$N-2 due to the shrinking of bands along the $\Gamma$-K-M path as is seen in \fig{fig02}(b-c). The peak here is due to optically active interband excitations through the Dirac point.
Consequently, the transition from $\pi$ to $\pi^*$ needs less energy, and the peak appears at a lower energy. In addition, a slightly enhancement in the optical conductivity is seen for BC$_6$N-2, which will be useful for applications in photovoltaic cells.

\subsection{Thermal Properties}

\begin{figure}[htb]
	\centering
	\includegraphics[width=0.5\textwidth]{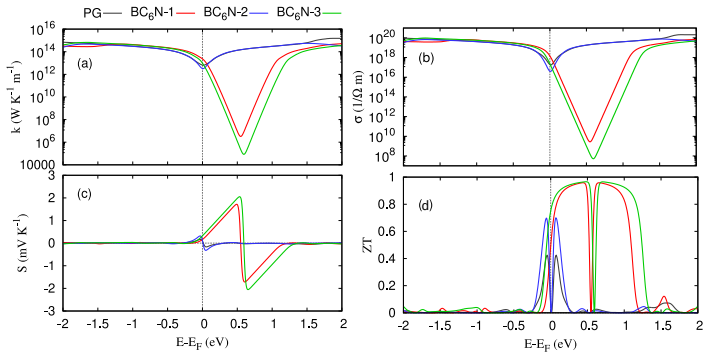}
	\caption{Electronic thermal conductivity, $k$ (a), electrical conductivity, $\sigma$ (b), Seebeck coefficient, $S$ (c), and figure of merit, ZT (d) versus
		energy are plotted PG (black), BC$_6$N-1 (red), BC$_6$N-2 (blue), and BC$_6$N-3 (green).}
	\label{fig11}
\end{figure}

In this section, thermal properties such as the electronic thermal conductivity (a), the electrical conductivity (b), the Seebeck coefficient (c), and the figure of merit (d) are investigated and displayed in in \fig{fig11}. The thermal calculations are carried out in the temperature range of $T = 20\text{-}150$~K where the electron contribution to the transport is dominant and the phonon participation can be neglected because the electron and lattice temperature are decoupled in this temperature range \cite{PhysRevB.87.241411, ABDULLAH2020126578}. BoltzTrap software can be thus used to investigate the thermal properties under the above condition \cite{Madsen2006}. The Boltztrap is based on the Boltzmann theory which is used to calculate the semiclassical transport coefficients. The code uses a mesh of band energies and is interfaced to the QE package \cite{RASHID2019102625}.

The thermal efficiency of a material or a device is characterized by the figure of merit, 
$ZT = (S^2 \, \sigma/k) \, T$, where $S$ is the Seebeck coefficient, $\sigma$ is the electrical conductivity, $T$ is a temperature, and $k$ is the electronic thermal conductivity.  
It has been shown that a material with high of $ZT$ should have high $S$, and $\sigma$, and low 
$k$ with a specificity temperature \cite{C7EE02007D}. Based on this, we can see that 
the BC$_6$N-1 and BC$_6$N-3 have the highest $ZT$ among all investigated structures in which 
the high symmetry breaking plays an important role. 
The high value of $ZT$ here refers to the relatively larger bandgap and asymmetry in the density of states of these two structures around the Fermi energy which give rise
to a lack of thermal electrons in the bandgap range. It thus leads to a decrease in the $k$ (a) and an increase in the $S$ (c) around the bandgap, which are the conditions to obtain the high value of $ZT$. The high value of $ZT$ of a material is useful to improve thermoelectric devices such as solar cell.
 
\section{Conclusions}\label{Sec:Conclusion}

To conclude, we have used density functional theory to study the structural, thermal, and optical properties of monolayer graphene doped with boron and nitrogen atoms where the influences of an attractive interaction between the dopant atoms are highlighted. We have assumed few atomic configurations forming different BC$_6$N structures based on the interaction between the B and N atoms.
We have found that the static dielectric constant is decreased, and the plasmon oscillation is reduced or vanishes in the presence of an attractive interaction between the B and N atoms compared to the pristine monolayer graphene. Furthermore, the refractive index indicating the transmittance behavior is enhanced in the case interactions between the B and N atoms in the low energy ranges while the reflectivity is decreased.
In addition, a high Seebeck coefficient and a high figure of merit have been found for BC$_6$N where the B and N atoms are doped at different atomic sites of the hexagon structure. 
We can thus confirm that the thermal and the optical properties of BC$_6$N can be improved by controlling the interaction between the dopant atoms which may be useful for thermoelectric and optoelectronic devises.



\begin{thebibliography}{10}
	
	\bibitem{doi:10.1063/1.4953172}
	Candela Mansilla~Wettstein, Franco~P. Bonafé, M.~Belén Oviedo, and
	Cristián~G. Sánchez.
	\newblock Optical properties of graphene nanoflakes: Shape matters.
	\newblock {\em The Journal of Chemical Physics}, 144(22):224305, 2016.
	
	\bibitem{C6CP07455C}
	Mehdi Shakourian-Fard and Ganesh Kamath.
	\newblock The effect of defect types on the electronic and optical properties
	of graphene nanoflakes physisorbed by ionic liquids.
	\newblock {\em Phys. Chem. Chem. Phys.}, 19:4383--4395, 2017.
	
	\bibitem{Wang206}
	Feng Wang, Yuanbo Zhang, Chuanshan Tian, Caglar Girit, Alex Zettl, Michael
	Crommie, and Y.~Ron Shen.
	\newblock Gate-variable optical transitions in graphene.
	\newblock {\em Science}, 320(5873):206--209, 2008.
	
	\bibitem{Nair1308}
	R.~R. Nair, P.~Blake, A.~N. Grigorenko, K.~S. Novoselov, T.~J. Booth,
	T.~Stauber, N.~M.~R. Peres, and A.~K. Geim.
	\newblock Fine structure constant defines visual transparency of graphene.
	\newblock {\em Science}, 320(5881):1308--1308, 2008.
	
	\bibitem{Fan2018}
	Yansong Fan, Chucai Guo, Zhihong Zhu, Wei Xu, Fan Wu, Xiaodong Yuan, and
	Shiqiao Qin.
	\newblock Monolayer-graphene-based broadband and wide-angle perfect absorption
	structures in the near infrared.
	\newblock {\em Scientific Reports}, 8(1):13709, 2018.
	
	\bibitem{Romagnoli2018}
	Marco Romagnoli, Vito Sorianello, Michele Midrio, Frank H.~L. Koppens, Cedric
	Huyghebaert, Daniel Neumaier, Paola Galli, Wolfgang Templ, Antonio D'Errico,
	and Andrea~C. Ferrari.
	\newblock Graphene-based integrated photonics for next-generation datacom and
	telecom.
	\newblock {\em Nature Reviews Materials}, 3(10):392--414, Oct 2018.
	
	\bibitem{Jia2012}
	Chuancheng Jia, Jiaolong Jiang, Lin Gan, and Xuefeng Guo.
	\newblock Direct optical characterization of graphene growth and domains on
	growth substrates.
	\newblock {\em Scientific Reports}, 2(1):707, Oct 2012.
	
	\bibitem{Ba2017}
	Kun Ba, Wei Jiang, Jingxin Cheng, Jingxian Bao, Ningning Xuan, Yangye Sun, Bing
	Liu, Aozhen Xie, Shiwei Wu, and Zhengzong Sun.
	\newblock Chemical and bandgap engineering in monolayer hexagonal boron
	nitride.
	\newblock {\em Scientific Reports}, 7(1):45584, Apr 2017.
	
	\bibitem{doi:10.1021/nl304351z}
	Aurélien Lherbier, Andrés~Rafael Botello-Méndez, and Jean-Christophe
	Charlier.
	\newblock Electronic and transport properties of unbalanced sublattice n-doping
	in graphene.
	\newblock {\em Nano Letters}, 13(4):1446--1450, 2013.
	\newblock PMID: 23477418.
	
	\bibitem{C5TA10599D}
	Stefano Agnoli and Marco Favaro.
	\newblock Doping graphene with boron: a review of synthesis methods{,}
	physicochemical characterization{,} and emerging applications.
	\newblock {\em J. Mater. Chem. A}, 4:5002--5025, 2016.
	
	\bibitem{Zhang2016}
	Jia Zhang, Chao Zhao, Na~Liu, Huanxi Zhang, Jingjing Liu, Yong~Qing Fu, Bin
	Guo, Zhenlong Wang, Shengbin Lei, and PingAn Hu.
	\newblock Tunable electronic properties of graphene through controlling bonding
	configurations of doped nitrogen atoms.
	\newblock {\em Scientific Reports}, 6(1):28330, Jun 2016.
	
	\bibitem{GOUDARZI2019130}
	M.~Goudarzi, S.S. Parhizgar, and J.~Beheshtian.
	\newblock Electronic and optical properties of vacancy and b, n, o and f doped
	graphene: Dft study.
	\newblock {\em Opto-Electronics Review}, 27(2):130--136, 2019.
	
	\bibitem{abdullah2021conversion}
	Nzar~Rauf Abdullah, Hunar~Omar Rashid, Chi-Shung Tang, Andrei Manolescu, and
	Vidar Gudmundsson.
	\newblock Conversion of the stacking orientation of bilayer graphene due
	to$\backslash$break the interaction of bn-dopants.
	\newblock {\em arXiv preprint arXiv:2101.00462}, 2021.
	
	\bibitem{Bu2020}
	Hongxia Bu, Haibin Zheng, Hongyu Zhang, Huimin Yuan, and Jingfen Zhao.
	\newblock Optical properties of a hexagonal c/bn framework with sp2 and sp3
	hybridized bonds.
	\newblock {\em Scientific Reports}, 10(1):6808, Apr 2020.
	
	\bibitem{ABDULLAH2021110095}
	Nzar~Rauf Abdullah, Hunar~Omar Rashid, Chi-Shung Tang, Andrei Manolescu, and
	Vidar Gudmundsson.
	\newblock Role of interlayer spacing on electronic, thermal and optical
	properties of bn-codoped bilayer graphene: Influence of the interlayer and
	the induced dipole-dipole interactions.
	\newblock {\em Journal of Physics and Chemistry of Solids}, 155:110095, 2021.
	
	\bibitem{MORTAZAVI2019733}
	Bohayra Mortazavi, Masoud Shahrokhi, Mostafa Raeisi, Xiaoying Zhuang, Luiz
	Felipe~C. Pereira, and Timon Rabczuk.
	\newblock Outstanding strength, optical characteristics and thermal
	conductivity of graphene-like bc3 and bc6n semiconductors.
	\newblock {\em Carbon}, 149:733--742, 2019.
	
	\bibitem{RANI201428}
	Pooja Rani, Girija~S. Dubey, and V.K. Jindal.
	\newblock Dft study of optical properties of pure and doped graphene.
	\newblock {\em Physica E: Low-dimensional Systems and Nanostructures},
	62:28--35, 2014.
	
	\bibitem{Giannozzi_2009}
	Paolo Giannozzi, Stefano Baroni, Nicola Bonini, Matteo Calandra, Roberto Car,
	Carlo Cavazzoni, Davide Ceresoli, Guido~L Chiarotti, Matteo Cococcioni,
	Ismaila Dabo, Andrea~Dal Corso, Stefano de~Gironcoli, Stefano Fabris, Guido
	Fratesi, Ralph Gebauer, Uwe Gerstmann, Christos Gougoussis, Anton Kokalj,
	Michele Lazzeri, Layla Martin-Samos, Nicola Marzari, Francesco Mauri,
	Riccardo Mazzarello, Stefano Paolini, Alfredo Pasquarello, Lorenzo Paulatto,
	Carlo Sbraccia, Sandro Scandolo, Gabriele Sclauzero, Ari~P Seitsonen,
	Alexander Smogunov, Paolo Umari, and Renata~M Wentzcovitch.
	\newblock {QUANTUM} {ESPRESSO}: a modular and open-source software project for
	quantum simulations of materials.
	\newblock {\em Journal of Physics: Condensed Matter}, 21(39):395502, sep 2009.
	
	\bibitem{giannozzi2017advanced}
	Paolo Giannozzi, Oliviero Andreussi, Thomas Brumme, Oana Bunau, M~Buongiorno
	Nardelli, Matteo Calandra, Roberto Car, Carlo Cavazzoni, Davide Ceresoli,
	Matteo Cococcioni, et~al.
	\newblock Advanced capabilities for materials modelling with quantum espresso.
	\newblock {\em Journal of Physics: Condensed Matter}, 29(46):465901, 2017.
	
	\bibitem{PhysRev.140.A1133}
	W.~Kohn and L.~J. Sham.
	\newblock Self-consistent equations including exchange and correlation effects.
	\newblock {\em Phys. Rev.}, 140:A1133--A1138, Nov 1965.
	
	\bibitem{Petersen2000}
	Max Petersen, Frank Wagner, Lars Hufnagel, Matthias Scheffler, Peter Blaha, and
	Karlheinz Schwarz.
	\newblock Improving the efficiency of fp-lapw calculations.
	\newblock {\em Computer Physics Communications}, 126(3):294--309, 2000.
	
	\bibitem{PhysRevLett.77.3865}
	John~P. Perdew, Kieron Burke, and Matthias Ernzerhof.
	\newblock Generalized gradient approximation made simple.
	\newblock {\em Phys. Rev. Lett.}, 77:3865--3868, Oct 1996.
	
	\bibitem{PhysRevB.33.8800}
	John~P. Perdew and Wang Yue.
	\newblock Accurate and simple density functional for the electronic exchange
	energy: Generalized gradient approximation.
	\newblock {\em Phys. Rev. B}, 33:8800--8802, Jun 1986.
	
	\bibitem{PhysRevB.23.5048}
	J.~P. Perdew and Alex Zunger.
	\newblock Self-interaction correction to density-functional approximations for
	many-electron systems.
	\newblock {\em Phys. Rev. B}, 23:5048--5079, May 1981.
	
	\bibitem{Mistrik2017}
	Jan Mistrik, Safa Kasap, Harry~E. Ruda, Cyril Koughia, and Jai Singh.
	\newblock {\em Optical Properties of Electronic Materials: Fundamentals and
		Characterization}, pages 1--1.
	\newblock Springer International Publishing, Cham, 2017.
	
	\bibitem{ABDULLAH2020103282}
	Nzar~Rauf Abdullah, Danyal~A. Abdalla, Taha~Y. Ahmed, Sarbast~W. Abdulqadr, and
	Hunar~Omar Rashid.
	\newblock Effect of bn dimers on the stability, electronic, and thermal
	properties of monolayer graphene.
	\newblock {\em Results in Physics}, 18:103282, 2020.
	
	\bibitem{ABDULLAH2020114556}
	Nzar~Rauf Abdullah, Hunar~Omar Rashid, Chi-Shung Tang, Andrei Manolescu, and
	Vidar Gudmundsson.
	\newblock Properties of bsi6n monolayers derived by first-principle
	computation.
	\newblock {\em Physica E: Low-dimensional Systems and Nanostructures}, page
	114556, 2020.
	
	\bibitem{doi:10.1063/1.4742063}
	Nabil Al-Aqtash, Khaldoun~M. Al-Tarawneh, Tarek Tawalbeh, and Igor Vasiliev.
	\newblock Ab initio study of the interactions between boron and nitrogen
	dopants in graphene.
	\newblock {\em Journal of Applied Physics}, 112(3):034304, 2012.
	
	\bibitem{ABDULLAH2020100740}
	Nzar~Rauf Abdullah, Hunar~Omar Rashid, Andrei Manolescu, and Vidar Gudmundsson.
	\newblock Interlayer interaction controlling the properties of ab- and
	aa-stacked bilayer graphene-like bc14n and si2c14.
	\newblock {\em Surfaces and Interfaces}, 21:100740, 2020.
	
	\bibitem{ABDULLAH2020126350}
	Nzar~Rauf Abdullah, Hunar~Omar Rashid, Mohammad~T. Kareem, Chi-Shung Tang,
	Andrei Manolescu, and Vidar Gudmundsson.
	\newblock Effects of bonded and non-bonded b/n codoping of graphene on its
	stability, interaction energy, electronic structure, and power factor.
	\newblock {\em Physics Letters A}, 384(12):126350, 2020.
	
	\bibitem{C2NR11728B}
	Xiaofeng Fan, Zexiang Shen, A.~Q. Liu, and Jer-Lai Kuo.
	\newblock Band gap opening of graphene by doping small boron nitride domains.
	\newblock {\em Nanoscale}, 4:2157--2165, 2012.
	
	\bibitem{ABDULLAH2021114644}
	Nzar~Rauf Abdullah, Mohammad~T. Kareem, Hunar~Omar Rashid, Andrei Manolescu,
	and Vidar Gudmundsson.
	\newblock Spin-polarised dft modeling of electronic, magnetic, thermal and
	optical properties of silicene doped with transition metals.
	\newblock {\em Physica E: Low-dimensional Systems and Nanostructures},
	129:114644, 2021.
	
	\bibitem{ABDULLAH2020126807}
	Nzar~Rauf Abdullah, Hunar~Omar Rashid, Chi-Shung Tang, Andrei Manolescu, and
	Vidar Gudmundsson.
	\newblock Modeling electronic, mechanical, optical and thermal properties of
	graphene-like bc6n materials: Role of prominent bn-bonds.
	\newblock {\em Physics Letters A}, 384(32):126807, 2020.
	
	\bibitem{doi:10.1021/jp2016616}
	Jun Zhu, Sumanta Bhandary, Biplab Sanyal, and Henrik Ottosson.
	\newblock Interpolation of atomically thin hexagonal boron nitride and
	graphene: Electronic structure and thermodynamic stability in terms of
	all-carbon conjugated paths and aromatic hexagons.
	\newblock {\em The Journal of Physical Chemistry C}, 115(20):10264--10271,
	2011.
	
	\bibitem{doi:10.1021/jp302293p}
	Chung-Huai Chang, Xiaofeng Fan, Lain-Jong Li, and Jer-Lai Kuo.
	\newblock Band gap tuning of graphene by adsorption of aromatic molecules.
	\newblock {\em The Journal of Physical Chemistry C}, 116(25):13788--13794,
	2012.
	
	\bibitem{PhysRevB.81.155413}
	V.~G. Kravets, A.~N. Grigorenko, R.~R. Nair, P.~Blake, S.~Anissimova, K.~S.
	Novoselov, and A.~K. Geim.
	\newblock Spectroscopic ellipsometry of graphene and an exciton-shifted van
	hove peak in absorption.
	\newblock {\em Phys. Rev. B}, 81:155413, Apr 2010.
	
	\bibitem{PhysRevB.77.233406}
	T.~Eberlein, U.~Bangert, R.~R. Nair, R.~Jones, M.~Gass, A.~L. Bleloch, K.~S.
	Novoselov, A.~Geim, and P.~R. Briddon.
	\newblock Plasmon spectroscopy of free-standing graphene films.
	\newblock {\em Phys. Rev. B}, 77:233406, Jun 2008.
	
	\bibitem{nano8110962}
	Bin Qiu, Xiuwen Zhao, Guichao Hu, Weiwei Yue, Junfeng Ren, and Xiaobo Yuan.
	\newblock Optical properties of graphene/mos2 heterostructure: First principles
	calculations.
	\newblock {\em Nanomaterials}, 8(11), 2018.
	
	\bibitem{NATH2014275}
	Palash Nath, Suman Chowdhury, D.~Sanyal, and Debnarayan Jana.
	\newblock Ab-initio calculation of electronic and optical properties of
	nitrogen and boron doped graphene nanosheet.
	\newblock {\em Carbon}, 73:275--282, 2014.
	
	\bibitem{PhysRevB.87.241411}
	S.~{Yi\ifmmode \breve{g}\else {\u g}\fi{}en}, V.~Tayari, J.~O. Island, J.~M.
	Porter, and A.~R. Champagne.
	\newblock Electronic thermal conductivity measurements in intrinsic graphene.
	\newblock {\em Phys. Rev. B}, 87:241411, Jun 2013.
	
	\bibitem{ABDULLAH2020126578}
	Nzar~Rauf Abdullah, Gullan~Ahmed Mohammed, Hunar~Omar Rashid, and Vidar
	Gudmundsson.
	\newblock Electronic, thermal, and optical properties of graphene like sicx
	structures: Significant effects of si atom configurations.
	\newblock {\em Physics Letters A}, 384(24):126578, 2020.
	
	\bibitem{Madsen2006}
	Georg K.~H. Madsen and David~J. Singh.
	\newblock Boltztrap. a code for calculating band-structure dependent
	quantities.
	\newblock {\em Computer Physics Communications}, 175(1):67--71, 2006.
	
	\bibitem{RASHID2019102625}
	Hunar~Omar Rashid, Nzar~Rauf Abdullah, and Vidar Gudmundsson.
	\newblock Silicon on a graphene nanosheet with triangle- and dot-shape:
	Electronic structure, specific heat, and thermal conductivity from
	first-principle calculations.
	\newblock {\em Results in Physics}, 15:102625, 2019.
	
	\bibitem{C7EE02007D}
	G.~Jeffrey Snyder and Alemayouh~H. Snyder.
	\newblock Figure of merit zt of a thermoelectric device defined from materials
	properties.
	\newblock {\em Energy Environ. Sci.}, 10:2280--2283, 2017.
	
\end{thebibliography}

\end{document}